\documentclass[A4,11pt]{article}
\usepackage{amsfonts}
\usepackage{amsmath}

\newtheorem{lemma}{Lemma}
\newtheorem{theorem}{Theorem}%[section]
\newenvironment{proof}%
{\begin{trivlist}\item[\hspace*{\labelsep}{\it Proof.\/}]}%
{\hfill$\Box$\end{trivlist}}

\usepackage{latexsym,graphicx}
\begin{document}
\title{\bf Improved Online  Hypercube Packing } %\thanks{Supported in part by }}

\author{Xin Han$^1$ \hspace{3mm} Deshi Ye$^2$ \hspace{3mm} Yong Zhou$^3$
%%$^3$\thanks{Supported in part by the DFG Project AL
%%464/4-1, Eu-Project APPOL II and NSFC (10231060).}
\\ {\small $^1$ School of Informatics, Kyoto University, Kyoto
606-8501, Japan} \\ {\small hanxin@kuis.kyoto-u.ac.jp}
\\ {\small $^2$ College of Computer Science, 
   Zhejiang University, Hangzhou, 310027, China}
\\ {\small  yedeshi@zju.edu.cn }
\\ {\small $^3$  Graduate School of Science, Hokkaido University, Sapporo, Japan}
\\ {\small zhou@castor.sci.hokudai.ac.jp}
%%\\ {\small Georges-K\"ohler-Allee 79, 79110 Freiburg, Germany}
%% \\ {\small $^3$ Department of Mathematics, Zhejiang University, China}
%% \\ {\small zgc@zju.edu.cn}
}
\date{}
\maketitle

\baselineskip 13.4pt

\begin{abstract}
 In this paper, we study online multidimensional bin packing problem
 when all items are hypercubes.
 %%We propose a new algorithm for hypercube packing,
 Based on the techniques in one dimensional bin packing
 algorithm Super Harmonic  by Seiden,
 we give a  framework for online  hypercube packing problem
 and obtain new upper bounds of asymptotic competitive ratios.
 For square packing, we get an upper bound of 2.1439,
which is better than 2.24437.
 For cube packing, we also give a new upper bound 2.6852 
which is better than 2.9421 by Epstein and van Stee.
 
\end{abstract}

\section{Introduction}
The classical one-dimensional 
Bin Packing is one of the oldest and most well-studied problems
in computer science \cite{CGJ97}, \cite{cw97}.
In the early 1970's it was one of the first combinatorial optimization
problems for which the idea of worst-case performance guarantees was
investigated. It was also in this domain that
the idea of proving lower bounds on the performance of online algorithm
was first developed.
 In this paper, we consider a generalization of the 
classical bin packing problem: hypercube  packing problem.

\paragraph{\bf Problem Definition.} Let $d \ge 1$ be an integer.
%%In the $d$-dimensional bin packing problem
We receive a sequence $\delta$ of   items $p_1,p_2,...,p_n$.
Each item $p$ is a $d$-dimensional hypercube and  has a fixed size,
 which is $s(p)\times \cdots \times s(p)$,
i.e., $s(p)$ is the size of $p$ in any dimension.
We have an infinite number of bins,
each of which is a $d$-dimensional unit hypercube.
Each item must be assigned to a position $(x_1(p),...,x_d(p))$ of some bin, 
where $0 \le x_i(p)$ and 
$x_i(p) + s(p) \le 1$ for $1 \le i \le d$.
Further, the positions must be assigned in such a way 
that no two items in the same bin overlap.
Note that for $d=1$ the problem reduces to  the 
classic bin packing problem.
In this paper, we study the {\em online} version of this problem,
i.e., each item must be assigned in turn,
without knowledge of the next items.

\paragraph {\bf Asymptotic competitive ratio.} 
To evaluate an online algorithms for  bin packing,
 we use the standard measure {\em Asymptotic competitive ratio}
 which is defined as follows.

Given an input list {\em L} and an online algorithm
$A$, we denote by $OPT(L)$ and $A(L)$, respectively, the cost 
(number of bins used) by an optimal (offline) algorithm and the cost
by online algorithm $A$ for packing list $L$.
The {\em asymptotic competitive ratio} $R_A^{\infty}$
of algorithm $A$ is defined by
\[
     R_A^{\infty} =\lim_{k \to \infty} \sup_{L}\{ A(L)/OPT(L)| OPT(L) = k\}.
\]

\paragraph{\bf Previous results.} 
 On the classic online bin packing,
 Johnson, Demers, Ullman, Garey and Graham \cite{JDUGG74}
 showed that the First Fit algorithm has the competitive ratio 1.7.
 Yao \cite{Yao80} gave an upper bound of 5/3.
 Lee and Lee \cite{LL85} showed the Harmonic algorithm has 
 the competitive ratio 1.69103 and improved it to 1.63597.
 Ramanan, Brown, Lee and Lee \cite{RBLL89} improved the upper bound to
 1.61217. Currently,
 the best known upper bound is 1.58889 by Seiden \cite{s02}.
 On the lower bounds,
 Yao \cite{Yao80} showed no online algorithm has performance
 ratio less that 1.5.
 Brown \cite{B79} and Liang \cite{L80} independently improved 
 this lower bound to 1.53635.
 The lower bound currently stands at 1.54014, due to van Vliet \cite{Vliet92}.

 On online  hypercube packing,
 Coppersmith and Raghavan \cite{CR89} showed an upper bound of
 43/16 = 2.6875 for  online square packing
 and an upper bound 6.25 for online cube packing.
 The upper bound for square packing was improved to 
 395/162 $<$ 2.43828 by Seiden and van Stee \cite{SS03}. 
 For online cube packing,
 Miyazawa and Wakabayashi \cite{MW03} showed an upper bound of 3.954.
 Epstein and van Stee \cite{ES04} gave an upper bound of 2.2697 
 for square packing
 and an upper bound of 2.9421 for online cube packing.
 By using a computer program, 
 the upper bound for square packing was improved to 2.24437 by 
  Epstein and van Stee \cite{ES05b}.
 They \cite{ES05b} also  gave lower bounds of 1.6406 and 1.6680
  for square packing and cube packing, respectively. 
 
\paragraph{\bf Our contributions.}
 When the Harmonic algorithm \cite{LL85} is extended into
 the online hypercube packing problem, 
 the items of sizes $1/2 + \epsilon, 1/3 + \epsilon, 1/4+\epsilon,
 \dots$ are still the crucial items related to
 the asymptotic competitive ratio, 
 where $\epsilon >0$ is sufficiently small.
 Using the techniques in one dimensional bin packing,
 Epstein and van Stee \cite{ES05b}  combined the items of size in
 $(1/2, 1-\Delta]$ with the items of size in $(1/3, \Delta]$
 and improved the Harmonic algorithm for hypercube packing,
 where $\Delta$ is a specified number in $(1/3, 0.385)$.
 In this paper, we do not only consider the combinatorial packing
 for the items in $(1/2, 1-\Delta]$ and $(1/3, \Delta]$,
 but also other crucial items.
 Based on the techniques in one dimensional bin packing
 algorithm Super Harmonic  by Seiden \cite{s02},
 we classify all the items into 17 groups and
 give a framework for online  hypercube packing.
 To analyse our algorithm,
 we give a weighting system consisting of  four weighting functions.
 By the weighting functions, we show that
 for square packing,  the asymptotic competitive ratio of our algorithm
 is at most 2.1439 which is better than  2.24437\cite{ES05b},
 for cube packing, the ratio is at most 2.6852, 
 which is also better than  2.9421\cite{ES05b}.

\noindent {\bf Definition}:
  If an item $p$  of size (side length) $s(p) \le 1/M$, 
 where $M$ is a fixed integer,
 then call $p$ {\em small}, otherwise {\em large}.

%% \section{Previous Research}
%%  We first introduce previous research on online hypercube packing
%%  that  is useful  in our algorithm.
\section{Online packing small items}
%%  When an itme $h$  of size $s(h) \le 1/M$, where $M \ge 5$ is a fixed integer,
%%  we call $p$ {\em small}.
 
 The following algorithm for  packing small items is from
 \cite{CV93}, \cite{ES05a}.
 The key ideas are below:
\begin{enumerate}
 \item Classify all {\em small} squares into $M$ groups.
       In detail, for an item $p$  of size $s(p)$, 
       we classify it into group $i$ such that 
      $2^ks(p) \in (1/(i+1),1/i]$, where $i \in \{M,...,2M-1\}$
      and $k$ is an integer.      
 \item Exclusively pack items of the same group into bins,
       i.e., each bin is used to pack items belonged to the same group.
       During packing, one bin may be partitioned into sub-bins.
\end{enumerate}

 \noindent {\bf Definition}: 
 An item is defined to be of type $i$ if it belongs to group $i$.
 A sub-bin which received an item is said to be {\em used}.
 A sub-bin which is not used and not cut into smaller sub-bins is called
 {\em empty}.
 A bin is called {\em active} if it can still receive items,
 otherwise {\em closed}.

 Given an item $p$ of type $i$, where $2^ks(p) \in (1/(i+1),1/i]$,
 {\em algorithm} AssignSmall($i$)  works as followings.
 
 %%\noindent{\bf AssignSmall($i$)}:
 \begin{enumerate}
  \item  If there is an empty sub-bin of size $1/(2^ki)$,
         then the item is simply packed there.
  \item  Else, in the current bin, if there is no empty sub-bin 
         of size $1/(2^ji)$ for $j < k$, then close the bin 
         and open a new bin and partition it into sub-bins of size $1/i$.
         If $k=0$ then pack the item in one of sub-bins of size $1/i$.
         Else goes to next step. 
  \item  Take an empty sub-bin of size  $1/(2^ji)$ for a maximum $j < k$.
         Partition it into $2^d$ identical sub-bins.
         If the resulting sub-bins are larger than $1/(2^ki)$,
         then take {\em one} of them and partition it in the same way. 
         This is done until sub-bins of size $1/(2^ki)$ are reached.
         Then the item is packed into one such sub-bin.
 \end{enumerate} 

%% The following results are from \cite{ES05a}.
 \begin{lemma} \label{lemma:small}
  In the above algorithm, 
  \vspace{-5pt}
  \begin{description}
   \item i) at any time, there are at most $M$ {\em active} bins. 
 \vspace{-5pt}
   \item ii) in each closed bin of type $i \ge M$,
    the occupied volume is at least $(i^d -1)/(i+1)^d \ge (M^d -1)/(M+1)^d$.
  \end{description}
 \end{lemma}
 So, roughly speaking, a small item with size $x$ takes
 at most  $\frac{(M+1)^d} {(M^d -1)}\times x^d$ bin.
  
\section{Algorithm $\mathcal{A}$ for online hypercube packing}
 The key points in our online algorithm are
 \begin{enumerate}
  \item divide all items into {\em small} and {\em large} groups.
  \item pack small items by algorithm AssignSmall,
        pack large items by an extended Super Harmonic algorithm.
 \end{enumerate}

Before giving our algorithm,
we first give some definitions and descriptions about the algorithm,
which are similar with the ones in \cite{s02}, but some definitions
are different from the ones in \cite{s02}.
 
\noindent{\bf Classification of large items}:
Given an integer $M \ge 11$,
let $t_1 = 1 > t_2 > \cdots>t_{N+1} = 1/M > t_{N+2}=0$,
where $N$ is a fixed integer.
We define the interval $I_j$ to be $(t_{j+1},t_j]$ for 
$j = 1,...,N+1$
and say a large item $p$ of size $s(p)$ has type $i$ 
if $s(p) \in I_i$.

\noindent{\bf Definition}: An item of size $s$  has type $\tau(s)$, where
\[
      \tau(s) = j \qquad    \Leftrightarrow  \qquad  s \in I_j.
\]

\noindent{\bf Parameters in algorithm $\mathcal{A}$}:
An instance of the algorithm is described by the following parameters:
integers $N$ and $K$; real numbers 
$1 = t_1 > t_2 > \cdots > t_N > t_{N+1} = 1/M$,
$\alpha_1,...,\alpha_{N} \in [0,1]$
and $0 = \Delta_0 < \Delta_1 < \cdots< \Delta_K< 1/2$,
and a function $\phi:\{1,...,N\} \mapsto \{0,...,K\}$.

Next, we give the operation of our algorithm,
essentially, which is quite similar with 
the Super Harmonic algorithm \cite{s02}.
Each {\em large} item of type $j$ is assigned a color,
 {\em red} or {\em blue}.
The algorithm uses two sets of counters, 
$e_1,...,e_{N}$ and $s_1,...,s_N$,
all of which are initially zero.
$s_i$ keeps track of the total number of type $i$ items.
$e_i$ is the number of type $i$ items which get colored red.
For $1 \le i \le N$,
the invariant $e_i = \lfloor \alpha_i s_i \rfloor$ is maintained,
 i.e. the percentage of type $i$ items colored red is 
 approximately $\alpha_i$.

 We first introduce  some parameters used in Super Harmonic algorithm,
 then give the corresponding ones for $d$-dimensional packing.
 In one dimensional packing,
 a bin  can be placed at most 
 $\beta_i = \lfloor 1/t_i\rfloor$ items
  with size $t_i$.
  After packing $\beta_i$ type $i$ items, 
  there is 
 $\delta_i = 1 - t_i \beta_i$  space left.
 The rest space can be used for red items.
%%  For example, when a type $4$ is placed in a bin,
%%  in every dimension of a bin, there is a rest space as least 
%%  $\delta_4 = p_1 \ge 1/3$,
%%  we would like to use this space to pack red items of type $6,7$ etc.
 However, we sometimes use less than $\delta_i$ in  a bin
 in order to simplify the algorithm and its analysis,
 i.e., 
 we  use  $\mathcal{D}=\{ \Delta_1,...,\Delta_K\}$ instead of  the set of
$\delta_i$, for all $i$.
  $\Delta_{\phi(i)}$ is the amount of space used to hold red items in
 a bin  which  holds blue items of type $i$.
  We  therefore require that $\phi$ satisfy $\Delta_{\phi(i)} \le \delta_i$.
 $\phi(i)=0$ indicates that no red items are accepted.
 To ensure that every red item potentially can be packed,
 we require that  $\alpha_i =0$ for all $i$ such that 
 $t_i > \Delta_K$, that is, there are no red items of type $i$.  
 Define $\gamma_i=0$ if $t_i > \delta_K$ and $\gamma_i = \max\{1, \lfloor \Delta_1/t_i\rfloor\}$, otherwise.
 This is the number of red item of type $i$ placed in a bin.
 
 In $d$-dimensional packing,
 we place $\beta_i^d$  blue items of type $i$
  into a bin and introduce a new parameter  $\theta_i$ instead of 
 $\gamma_i$.
 Let
 \[
\theta_i  = \beta_i^d 
          -(\beta_i - \gamma_i)^d.
\] 
 This is the number of red items of type $i$ that the algorithm
 places together in  a bin.
 In details,
 if $t_i > \Delta_K$, then  $\theta_i = 0$, i.e., we do not pack
 type $i$ items as red items.
 So, in this case, we require $\alpha_i = 0$.
 Else if $t_i \le \Delta_1$, then $\theta_i  = \beta_i^d 
          -(\beta_i - \lfloor \Delta_1/t_i\rfloor)^d$.
 If $\Delta_1 < t_i \le \Delta_K$, we set 
 $\theta_i  = \beta_i^d 
          -(\beta_i - 1)^d$.

%%  Define $\gamma_i = 0$ if $t_i > \Delta_K$ and 
%%  $\gamma_i = \max\{1, \lfloor \Delta_1/t_i\rfloor\}$ otherwise.
 
%%  And  if possible, we would like to  pack red items into a bin with 
%% $\beta_i^d$ blue items of type $i$  when $\delta_i \ne 0$. 

%%  Define 
%% \[
%%   \varphi(i) = \min\{j |  t_i \le \Delta_j, 1 \le j \le K \}.
%% \]
%% Intuitively, $\varphi(i)$ is the smallest index 
%% to indicates that a red item of type $i$ can be packed along with
%% blue items of type $j$ in  a bin,
%%  where $\varphi(i) \le j \le K $.

 Here, we illustrate the structure of a bin for $d=2$.
\begin{figure}[htbp]
  \begin{center}
  \includegraphics[scale=0.7]{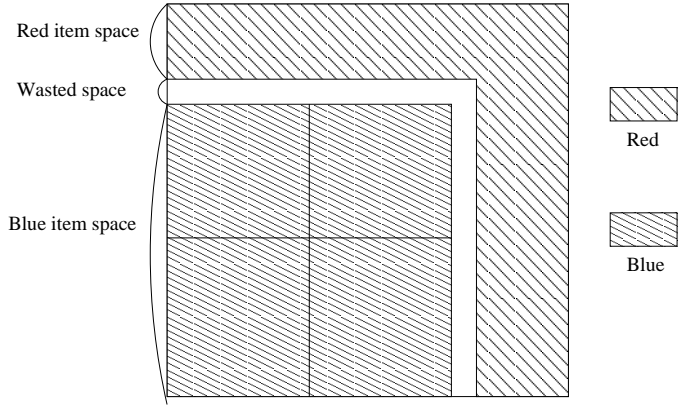}
  \caption{ If the bin is a (i, j) or (i, ?) bin,
            the amount of area for blue items is $(t_i \beta_i)^2$.
            The amount of area left is $1-(t_i \beta_i)^2$. 
            The amount of this area actually used for red items
            is $1 - (1-\Delta_{\phi(i)})^2$,
            where $\Delta_{\phi(i)} \le \delta_i = 1-t_i \beta_i $.}
  \label{fig:redblue}
  \end{center}
\end{figure}

%% We now specify the bin groups. They are named:
 {\bf Naming bins:}  Bins are named as follows:
 \begin{displaymath}
  \begin{array}{l}
    \{i|\phi_i = 0, 1 \le i \le N,\} \\
 
   \{(i,?)|\phi_i \ne 0, 1 \le i \le N,\} \\
 
   \{(?,j)|\alpha_j \ne 0, 1 \le j \le N,\} \\
 
    \{(i,j)|\phi_i \ne 0, \alpha_j \ne 0, \gamma_j t_j \le \Delta_{\phi(i)},
1 \le i,j \le N\}. 
  \end{array}
 \end{displaymath}
 We call these groups {\em monochromatic}, {\em indeterminate blue},
 {\em indeterminate red} and {\em bichromatic}, respectively.
And we call the monochromatic and bichromatic groups {\em final} groups.

The monochromatic group $i$ contains bins that hold only blue items of type 
$i$. There is only one open bin in each of these groups; 
this bin has fewer than $\beta_i^d$ items. 
The closed bins all contain $\beta_i^d$ items.

The bichromatic group $(i,j)$ contains bins that contain blue items of type
$i$ along with red items of type $j$. A closed bin in this group contains
$\beta_i^d$ type $i$ items and $\theta_j$ type $j$ items. There are at most
three open bins.

The indeterminate blue group $(i, ?)$ contains bins that hold only blue items
of type $i$. These bins are all open, but only one has fewer than
$\beta_i^d$ items. 

The indeterminate red group $( ?, j)$ contains bins that hold only red items
of type $j$. These bins are all open, but only one has fewer than
$\theta_j$ items. 

Essentially, the algorithm tries to minimize the number of indeterminate bins,
while maintaining all the aforementioned invariants.
That is, we try to place red and blue items together whenever possible;
when this is not possible we place them in indeterminate bins in hope that
they can later be so combined.

 \noindent{\bf Algorithm $\mathcal{A}$}: 
A formal description of algorithm $\mathcal{A}$ is given as blow:
 \begin{description}
  \item Initialize  $e_i \gets 0$ and  $s_i \gets 0$ for $1 \le i  \le M+1$.
  \item For a small item $p$, call algorithm AssignSmall.
  \item For a large item $p$:
   \begin{description}
   \item $i \gets \tau(p)$, \quad $s_i \gets s_i +1$.
   \item If $e_i < \lfloor \alpha_i s_i \rfloor$:
    \begin{description}
    \item  $e_i \gets e_i +1$.
    \item  Color $p$ red.
    \item If there is an open bin in group $(?,i)$ with
          fewer than $\theta_i$ type $i$ items, then pack $p$ in
          this bin.
     \item If there is an open bin in group $(j,i)$  with
          fewer than $\theta_i$ type $i$ items, then pack $p$ in
          this bin.
     \item Else if there is some bin in group $(j,?)$  such that
            $\Delta_{\phi(j)} \ge \gamma_i t_i$ then place
           $p$ in it and change the group of this bin to $(j,i)$.
      \item Otherwise, open a new group $(?, i)$ bin and place $p$ in it.
    \end{description}
   \item Else:
    \begin{description}
    \item Color $p$ blue.
    \item If $\phi(i) = 0 $:
     \begin{description}
      \item If there is an open bin in group $i$ with fewer than $\beta_i^d$
            items, then place $p$ in it.
      \item Otherwise, open a new group $i$ bin and pack $p$ there.
      \end{description}
    \item Else: 
    \begin{description}
     \item If, for any $j$, there is an open bin $(i,j)$ with fewer 
           than $\beta_i^d$ items, then place $p$ in this bin.

     \item Else,  if there is some bin in group $(i, ?)$ with fewer
           than $\beta_i^d$ items, then   place $p$ in this bin.
     \item Else, if there is some bin in group $(?,j)$ such that
           $\Delta_{\phi(i)} \ge \gamma_j t_j$  then pack $p$ in it 
           and change the group of this bin  to $(i,j)$.
      \item Otherwise, open a new group $(i,?)$ bin and pack $p$ there.
       \end{description}
     \end{description}
    \end{description}
 \end{description}
 
\baselineskip 13.4pt

\section{The analyses for square and cube packing }
 In this section, we fix the parameters in the framework given 
 in the last section for square packing and cube packing respectively. 
 Then we analyse the competitive ratios by 
 a corresponding weighting system consisting
 of four weighting functions.

\subsection{An instance of algorithm $\mathcal{A}$}
Let $M=11$, i.e.,  a {\em small} item has its side length as most $1/11$.
And the parameters in $\mathcal{A}$ are given in the following tables.
First we classify all the items into  17 groups by fixing the values of 
$t_i$, where $1 \le i \le 18$.
Then we calculate the number of blue type $i$ in a bin, $\beta_i^{d}$.
Finally, we define the set  $\mathcal{D}=\{ \Delta_1,...,\Delta_K\}$ 
and the function $\phi(i)$, 
which are related to how many red items $\theta_i^{d}$ can be accepted 
in a bin, where $K=4$.
Note that $\alpha_i$ which is the percentage of type $i$ items colored red
will be given later. For square packing, we use a set of $\alpha_i$.
While for cube packing, we use another set of $\alpha_i$.

%%\begin{table}[htbp]
 \begin{displaymath} 
 \begin{array}{|c|c|c|c|c|c|}
  \hline
    i & (t_{i+1},t_i] &  \beta_i &  \delta_i & \phi(i) & \gamma_i \\
  \hline
   1  & (0.7, 1]    & 1  &  0      &   0  & 0    \\
 %% \hline
   2  & (0.65, 0.7] & 1  &  0.3    &   2  &   0 \\
 %%  \hline
   3  & (0.60, 0.65] & 1  &  0.35  &  3   &  0  \\
%%  \hline
  4   & (0.5,  0.60] & 1  &  0.4   &  4   &  0  \\
%% \hline
  5   & (0.4, 0.5 ] & 2   &  0     & 0    &  0 \\
%%  \hline
 6    & (0.35, 0.4] & 2   &  0.2   & 1    &  1  \\
%%  \hline
 7    & (1/3,  0.35] & 2  &  0.3   & 2    &  1   \\
%% \hline
 8    & (0.30, 1/3]  & 3  &  0     & 0    &  0  \\
%%\hline
 9    & (1/4, 0.30]  & 3  &  0.1   & 0    &  1 \\
%%\hline
 10   & (1/5, 1/4 ]  & 4  &  0     & 0    &  1 \\
%%%\hline
 11   & (1/6, 1/5 ]  & 5  &  0     & 0    &  1  \\
%%\hline
 12   & (1/7,  1/6]  & 6  &  0     & 0    &  1  \\
 13   & (1/8,  1/7]  & 7  &  0     & 0    &  1  \\
 14   & (1/9,  1/8]  & 8  &  0     & 0    &  1  \\
 15   & (0.1,  1/9]  & 9  &  0     & 0    &  1   \\
 16   & (1/11, 0.1]  & 10 &  0     & 0    &  2  \\
 17   & (0,   1/11]  & * &  *     & *     &  *  \\
\hline
 \end{array}
\hspace{3mm}
\begin{array}{|c|c|c|}
 \hline
  j = \phi(i) & \Delta_j  &  \textrm{Red items accepted} \\
 \hline
  1 &  0.20    &  11..16   \\
  2 &  0.30    &  9..16   \\
  3 &  0.35    &  7, 9..16   \\
  4 &  0.40    &  6..7, 9..16   \\

 \hline
\end{array} 
 \end{displaymath}

\noindent{\bf Observation:}  By the above tables,
in any dimension of  a $(4,?)$ bin,  
the distance  between the type 4 item and the opposite edge (face) of
the bin is at least    $\Delta_4 = 0.4$, 
since we pack a type 4 item in a corner of a bin.
So, all red items with size at most 0.4 can be packed in $(4,?)$ bins.
In the same ways,
all red items with  size at most 0.35 can be packed in $(4,?)$ and $(3,?)$ bins,
all red items with size at most 0.30 can be packed in
$(4,?)$, $(3,?)$, $(7,?)$ and $(2,?)$ bins,
all red items with size at most 0.2 can be packed in
$(4,?)$, $(3,?)$, $(7,?)$, $(2,?)$, $(6,?)$ bins.

 Next we define the weight function $W(p)$ for a given item $p$ with size $x$.
 Roughly speaking, a weight of an item is the maximal portion of a bin
 that it can occupy.
 Given a small item $p$ with size $x$,  by Lemma \ref{lemma:small},
 it  occupies a $\frac{x^d (11+1)^d}{11^d -1}$ bin.
 So, we define 
 \[
W(p) = \frac{x^d (11+1)^d}{11^d -1}.
 \]

 Given a large item  $p$, we consider four cases to 
 define its weight.
 Let $R_i$ and $B_i$ be the number of bins containing blue items
 of type $i$ and red items of type $i$, respectively.
 Let $E$ be the number of indeterminate red group bins, 
 i.e., some bins like $(?,i)$.
 If $E > 0$ then there are some $(?, j)$ bins.
 Let 
\[
e = \min\{j| (?,j)\},
\] 
 which is the type of 
 the smallest red item in an indeterminate red group bin.
 Let $\mathcal{A}(L)$ be the number of bins used by $\mathcal{A}$.

  \noindent {\bf Case 1}: $E=0$, i.e., no indeterminate red bins.
  Then every red item is packed with one or more blue items.
  Therefore
  \[
    \mathcal{A}(L) \le  \mathcal{A}(L_s) + \sum_{i} B_i,
  \]
  where $\mathcal{A}(L_s)$ is the number of bins for  small items.
  Since there are a constant number of active bins and
  every closed blue bin $(i)$ or $(i,*)$ contains $\frac{1}{\beta_i^d}$ items,
  we define the weighting function as below:
  \begin{displaymath}
  W_{1,1}(p)  = \frac{1-\alpha_i}{\beta_i^d}  \qquad \textrm{if } x \in I_i,  \textrm{ for } i=1..16. 
 \end{displaymath}

%%   \begin{displaymath}
%%   \begin{array}{ll}
%%   W_1(p)  = &   \left\{   
%% %%                                  { \setlength \arrayrowsep{1mm}
%%                                     \begin{array}{ll}
%%                 \frac{1}{\beta_i^d}     &  \textrm{if } x \in I_i, \textrm{ for } i=1..5,8.  \\
%%  \vspace{4pt}
%%  \frac{1-\alpha_i}{\beta_i^d}  & \textrm{if } x \in I_i,  \textrm{ for } i=6,7,9..16. 
%%                                     \end{array}
%%                                     \right. 
%%   \end{array}
%%  \end{displaymath}

 \noindent {\bf Case 2}: $E>0$ and $e = 6$. Then 
 there are some bins $(?,6)$ and  no other bins $(?,j)$ bins, 
 where $j > 6$. Since a type 4 item can be packed
 into a bin  $(?,6)$, it is impossible to have bins $(4, ?)$.
 If we count  all $(4,j)$ bins as  red bins, 
 then 
  \[
    \mathcal{A}(L) \le  \mathcal{A}(L_s) + \sum_{i=1..3,5,8} B_i
     + \sum_{i=6,7,9..16} (R_i + B_i).
  \]
 Else we count  all $(4,j)$ bins as blue  bins then
  
  \[
    \mathcal{A}(L) \le  \mathcal{A}(L_s) + \sum_{i=1..16} B_i
     +  R_6.
  \]
   Since there are a constant number of active bins and
  every closed blue bin $(i)$ or $(i,*)$ contains $\frac{1}{\beta_i^d}$ items,
  every closed red bin $(j,i)$ or $(?,i)$ contains $\frac{1}{\theta_i}$ items,
  we define the weighting functions for two subcases as below:
 \begin{displaymath}
  \begin{array}{ll}
  W_{2,1}(p)  =  &   \left\{ 
                                    \begin{array}{ll}
                 \frac{1}{\beta_i^d}      &  \textrm{if } x \in I_i,
 \textrm{ for } i= 1,2,3,5,8. \\      
\vspace{4pt}                  
                                     0     &  \textrm{if } x \in I_4. \\
%%  \vspace{4pt}
%%  \frac{1-\alpha_i}{\beta_i^d} + \frac{\alpha_i}{\theta_i}  & \textrm{if } x \in I_i, \textrm{ for } i= 6.\\
\vspace{4pt}
 \frac{1-\alpha_i}{\beta_i^d} + \frac{\alpha_i}{\theta_i}  & \textrm{if } x \in I_i,  \textrm{ for } i=6,7,9..16. 
                                    \end{array}
                             \right. 
  \end{array}
 \end{displaymath}
  and 
   \begin{displaymath}
  \begin{array}{ll}
  W_{2,2}(p)  =  &   \left\{ 
                                    \begin{array}{ll}
               \frac{1-\alpha_i}{\beta_i^d}        &  \textrm{if } x \in I_i,
 \textrm{ for } i= 1..5,7..16. \\      
%% \vspace{4pt}                  
%%                                      1    &  \textrm{if } x \in I_4. \\
 \vspace{4pt}
 \frac{1-\alpha_i}{\beta_i^d} + \frac{\alpha_i}{\theta_i}  & \textrm{if } x \in I_i, \textrm{ for } i= 6.
%% \vspace{4pt}
%%  \frac{1-\alpha_i}{\beta_i^d}  & \textrm{if } x \in I_i,  \textrm{ for } i=7,9..16. 
                                    \end{array}
                             \right. 
  \end{array}
 \end{displaymath}

 \noindent {\bf Case 3}: $E>0$ and $e = 7$. Then 
 there are some bins  $(?,7)$ and no other bins $(?,j)$, 
 where $j > 7$. Since a type 4  or a type 3 item can be packed
 into a bin  $(?,7)$, it is impossible to have bins $(4, ?)$ and $(3, ?)$.
 If we count  all $(4,j)$ and $(3, j)$ bins as  red bins, 
 then 
  \[
    \mathcal{A}(L) \le  \mathcal{A}(L_s) + \sum_{i=1,2,5,8} B_i
     + \sum_{i=6,7,9..16} (R_i + B_i).
  \]
 Else we count  all $(4,j)$ and $(3, j)$   bins as blue  bins then
  
  \[
    \mathcal{A}(L) \le  \mathcal{A}(L_s) + \sum_{i=1..16} B_i
     +  R_6 + R_7.
  \]
%%    Since there are a constant number of active bins and
%%   every closed blue bin $(i)$ or $(i,*)$ contains $\frac{1}{\beta_i^d}$ items,
%%   every closed red bin $(j,i)$ or $(?,i)$ contains $\frac{1}{\theta_i}$ items,
  We define the weighting functions for two subcases as below:
  \begin{displaymath}
  \begin{array}{ll}
  W_{3,1}(p)  =  &   \left\{ 
                                    \begin{array}{ll}
                 \frac{1}{\beta_i^d}      &  \textrm{if } x \in I_i,
 \textrm{ for } i= 1,2,5,8. \\      
\vspace{4pt}                  
                                     0     &  \textrm{if } x \in I_3, I_4. \\
%%  \vspace{4pt}
%%  \frac{1-\alpha_i}{\beta_i^d} + \frac{\alpha_i}{\theta_i}  & \textrm{if } x \in I_i, \textrm{ for } i= 6.\\
\vspace{4pt}
 \frac{1-\alpha_i}{\beta_i^d} + \frac{\alpha_i}{\theta_i}  & \textrm{if } x \in I_i,  \textrm{ for } i=6,7,9..16. 
                                    \end{array}
                             \right. 
  \end{array}
 \end{displaymath}
  and 
   \begin{displaymath}
  \begin{array}{ll}
  W_{3,2}(p)  =  &   \left\{ 
                                    \begin{array}{ll}
               \frac{1-\alpha_i}{\beta_i^d}        &  \textrm{if } x \in I_i,
 \textrm{ for } i= 1..5,8..16. \\      
%% \vspace{4pt}                  
%%                                      1    &  \textrm{if } x \in I_4. \\
 \vspace{4pt}
 \frac{1-\alpha_i}{\beta_i^d} + \frac{\alpha_i}{\theta_i}  & \textrm{if } x \in I_i, \textrm{ for } i= 6, 7.
%% \vspace{4pt}
%%  \frac{1-\alpha_i}{\beta_i^d}  & \textrm{if } x \in I_i,  \textrm{ for } i=7,9..16. 
                                    \end{array}
                             \right. 
  \end{array}
 \end{displaymath}

%%   \begin{displaymath}
%%   \begin{array}{ll}
%%   W_{3,1}(p)  = &   \left\{ 
%%                                     \begin{array}{ll}
%%        \frac{1}{\beta_i^d}   &  \textrm{if } x \in I_i, \textrm{ for } i=1,2,5,8.  \\      
%% \vspace{4pt}                  
%%                                      0     &  \textrm{if } x \in I_i, \textrm{ for } i=3,4. \\
%%  \vspace{4pt}
%%  \frac{1-\alpha_i}{\beta_i^d} + \frac{\alpha_i}{\theta_i}  & \textrm{if } x \in  I_i, \textrm{ for } i=6,7.\\
%% \vspace{4pt}
%%  \frac{1-\alpha_i}{\beta_i^d} + \frac{\alpha_i}{\theta_i}   & \textrm{if } x \in I_i,  \textrm{ for } i=9..16. 
%%                                     \end{array}
%%                              \right. 
%%   \end{array}
%%  \end{displaymath}
%%  and 
%%   \begin{displaymath}
%%   \begin{array}{ll}
%%   W_{3,2}(p)  =  &   \left\{ 
%%                                     \begin{array}{ll}
%%        \frac{1}{\beta_i^d}   &  \textrm{if } x \in I_i, \textrm{ for } i=1,2,5,8.  \\      
%% \vspace{4pt}                  
%%                                      1     &  \textrm{if } x \in I_i, \textrm{ for } i=3,4. \\
%%  \vspace{4pt}
%%  \frac{1-\alpha_i}{\beta_i^d} + \frac{\alpha_i}{\theta_i}  & \textrm{if } x \in  I_i, \textrm{ for } i=6,7.\\
%% \vspace{4pt}
%%  \frac{1-\alpha_i}{\beta_i^d}  & \textrm{if } x \in I_i,  \textrm{ for } i=9..16. 
%%                                     \end{array}
%%                              \right. 
%%   \end{array}
%%  \end{displaymath}

\noindent {\bf Case 4}: $E>0$ and $e \ge 9$. Then 
 there are some bins  $(?,9)$. Since a type 2,3,4,7 item can be packed
 into a bin  $(?,9)$, it is impossible to have bins $(2, ?)$, $(3, ?)$, 
 $(4, ?)$,  $(7, ?)$.
 If we count  these bins  $(2, j)$, $(3, j)$, 
  $(4, j)$,  $(7, j)$ as  red bins, 
 then 
  \[
    \mathcal{A}(L) \le  \mathcal{A}(L_s) + \sum_{i=1,5,8} B_i
     + \sum_{i=6,9..16} (R_i + B_i) + R_7.
  \]
%%  Else we count  all $(4,j)$ and $(3, j)$   bins as blue  bins then
  
%%   \[
%%     \mathcal{A}(L) \le  \mathcal{A}(L_s) + \sum_{i=1..16} B_i
%%      +  R_6 + R_7.
%%   \]
 We define the weighting function as below:
 \begin{displaymath}
  \begin{array}{ll}
  W_{4,1}(p)  = & \left\{ 
                                    \begin{array}{ll}
                \frac{1}{\beta_i^d}      &  \textrm{if } x \in I_i, \textrm{ for } i=1,5,8.  \\      
\vspace{4pt}                  
                                     0     &  \textrm{if } x \in  I_2,I_3, I_4 \\
 \vspace{4pt}
 \frac{1-\alpha_i}{\beta_i^d} + \frac{\alpha_i}{\theta_i}  & \textrm{if } x \in I_i,  \textrm{ for } i=6,9..16 \\
 \vspace{4pt}
  \frac{\alpha_i}{\theta_i}  & \textrm{if } x \in I_i, \textrm{ for } i=7. 
%%  \vspace{4pt}
%%  \frac{1-\alpha_i}{\beta_i^d}  & \textrm{if } x \in I_i,  \textrm{ for } i=11..16.
                                    \end{array}
                             \right. 
 \end{array}
 \end{displaymath}

\noindent {\bf Definition}:
 A set of items $X$ is a feasible set 
if all items in it can be packed into a bin.
And,
\[
 W_{i,j}(X) = \sum_{p \in X} W_{i,j}(p) . 
\]

 Over all feasible sets $X$,  let 
\[
  W_i(X) = \min\{W_{i,j}(X)\}, \textrm { j = 1 or 2},
\]
and  define 
\[
\mathcal{P}(W) = \max\{ W_i(X)\} \textrm { for all } i.
\]

We defined four sets of weighting functions for all items. 
This is a weighting system,
which is a special case of general weighting system defined in \cite{s02}.
So, the following lemma follows directly from \cite{s02}.
\begin{lemma}
 \label{lemma:upper}
  The asymptotic performance ratio of $\mathcal{A}$  is upper bounded by $\mathcal{P}(W)$. 
\end{lemma}

\subsection{Upper bounds for square and cube packing}
In this subsection, 
we fix the parameters $\alpha_i$ for square packing
and cube packing respectively, 
and get the upper bounds of the asymptotic competitive ratios.

\noindent{\bf Definition} Let $m_i \ge 0$ be the number of 
 type $i$ items in a feasible set $X$.
 Given an item $p$ with size $x$,
 define an efficient function  $E_{i,j}(p)$ as $W_{i,j}(p)/x^d$.

\begin{theorem}
 The asymptotic performance ratio of $\mathcal{A}$ for square packing is at most 2.1439.
\end{theorem}
\begin{proof}
 For square packing, we set parameters $\alpha_i$  according to
 the following table.
 \begin{displaymath}
 \begin{array}{|c|l|l|l|l|l|l|l|l|l|l|l|l|l|}
 \hline
 i           & 1-4 & 5 & 6 & 7 & 8 & 9 & 10 & 11 & 12& 13& 14 &15 & 16 \\
 \hline
 \alpha_i  & 0 & 0& 0.12 & 0.2 & 0 & 0.2546 & 0.2096 & 0.15& 0.1& 0.1 &0.1& 0.1 & 0.05 \\
 \hline
 \theta_i & 0 &0 & 3 &  3 & 0 & 5 & 7 & 9 &11& 13 & 15 & 17 & 36    \\
 \hline
 \beta_i^2& 1 & 4 & 4 &4 & 9& 9 & 16& 25 & 36& 49& 64 & 81 & 100 \\
 \hline
 \end{array}
\end{displaymath}

 Based on the values in the followint two tables,
 we calculate  the upper bound of $\mathcal{P}(W) = \max\{W_i(X)\}$.
 \begin{displaymath}
 \begin{array}{|c|l|l|l|l|l|l|l|}
 \hline
  i & (t_{i+1},t_i] & W_{1,1}(p) & E_{1,1}(p)  & W_{2,1}(p) & E_{2,1}(p) &W_{2,2}(p) & E_{2,2}(p)  \\
 \hline
      1 &(0.7,1]   & 1     & 2.05   &   1     & 2.05   & 1     & 2.05     \\
      2 &(0.65,0.7]& 1     & 2.37   &   1     & 2.37   & 1     & 2.37     \\
      3 &(0.6,0.65]&  1     & 2.7778 &   1    & 2.7778&   1     & 2.7778   \\
      4 &(0.5,0.6] & 1     & 4      & 0       & 0      & 1     & 4         \\
      5 &(0.4,0.5]& 1/4   & 1.5625 & 1/4     & 1.5625 & 1/4   & 1.5625     \\
      6 & (0.35,0.4]& 0.22  & 1.8   & 0.26    & 2.123   & 0.26  & 2.123    \\
      7 &(1/3,0.35]& 0.2   & 1.8      & 0.8/3   & 2.4   &  0.2     & 1.8   \\
      8 &(0.3,1/3]&  1/9  & 1.235 &  1/9     & 1.235  &  1/9   & 1.235     \\
      9 &(1/4,0.3]& 0.0829 & 1.327 & 0.1338 & 2.141 & 0.0829 & 1.327  \\
   10..17 &(0,1/4]& 1.235x^2& 1.235  &1.99x^2 & 1.99   &1.235x^2 & 1.235    \\
 %%     11 &(1/6,1/5]& 0.034& 1.224   & 0.05067 & 1.824 & 0.034 & 1.224 \\
%%      12 &(1/7,1/6]& 0.025 & 1.225  & 0.03410  & 1.6705& 0.025 & 1.225 \\      
%%      13 &(1/8,1/7]& 0.01837 & 1.1756 & 0.02606  & 1.6679  & 0.01837 & 1.1756 \\  
%%      14 &(1/9,1/8]& 0.9/64 &1.2  & 0.02073  & 1.6791 & 0.9/64 &1.2  \\   
%%      15 &(0.1,1/9]& 0.1/9& 1.2   & 0.017   & 1.7   &  0.1/9  & 1.2  \\        
%%      16 &(1/11,0.1]& 0.0095 & 1.2  & 0.01089  & 1.3176  & 0.0095 & 1.2  \\  
%%      17 &(0,1/11]& 1.2x^2 & 1.2  &1.2x^2   & 1.2     &1.2x^2  & 1.2 \\   
 \hline
 \end{array}
 \end{displaymath}

%%  Based on the number of items $m_i$.
%%  we have 5 cases for 
%%  estimating the upper bound of $\mathcal{P}(W) = \max\{W_i(X)\}$.

\paragraph{ Case 1:} $ W_1(X) \le   2.1439$.
 
 \noindent If $m_2 + m_3 + m_4 = 0$, i.e., no  type $2,3,4$ items in $X$,
 then 
 \[
  W_1(X) = \sum_{p \in X} E_{1,1}(p) s(p)^2
   \le 2.05    \sum_{p \in X} s(p)^2 \le 2.05.
 \]  
 Else $m_2 + m_3 + m_4 = 1$. Then
 $m_5 + m_6 + m_7 \le 3$ and $m_6 + m_7 + m_9 \le 5$,
\[
  \begin{array}{lll}
    W_1(X) & \le &  1+  m_5/4 + 0.22 m_6 + 0.2m_7 + 0.0829m_9 
               + 1.235(1- \sum_{i=2}^{7}t_{i+1}^2 m_i -m_9/16) \\
                  & < &  2.1439.
   \end{array}
 \]
 The last inequality follows from $m_4=1$, $m_6=3$ and $m_9=2$.

\paragraph{ Case 2:} $ W_2(X) \le   2.134$.

 \noindent If $m_2 + m_3 + m_4 = 0$, i.e., no  type $2,3,4$ items in $X$,
 then 
 \[
  W_2(X) = \min\{ W_{2,1}(X), W_{2,2}(X)\} \le  W_{2,2}(X) \le  2.123.
 \]  
 Else $m_2 = 1$. Then no type $1,3,4,5,6$ items  in $X$.
 \[
  W_2(X) = W_{2,2}(X) \le 1 + 1.8 (1 - 0.65^2)  =  2.0395.
 \]  
 Else  $m_3 = 1$. Then no  type $1,2,4,5$ items in $X$ and 
 $ m_6 + m_7 \le 3$ and $m_6 + m_7 + m_9 \le 5$,
\[
  \begin{array}{lll}
    W_2(X) =  W_{2,2}(X) &\le& 1+ 0.26 m_6 + 0.2m_7 + 0.0829m_9 \\
                  & & + 1.235(1-0.6^2 -0.35^2m_6-m_7/9 -m_9/16) \\
                  & < &  2.134.
   \end{array}
 \]
The last inequality follows from $m_6=3$ and $m_9=2$.

\noindent Else $m_4 =1$. Then no type $1,2,3$ items  in $X$.
  \[
  W_2(X) \le W_{2,1}(X) \le 0 + 2.4 (1 - 0.5^2)  =  1.8.
 \]  

 \begin{displaymath}
 \begin{array}{|c|l|l|l|l|l|l|l|}
 \hline
  i & (t_{i+1},t_i] & W_{3,1}(p) & E_{3,1}(p)  & W_{3,2}(p) & E_{3,2}(p)  & W_{4,1}(p) & E_{4,1}(p) \\
 \hline
      1 &(0.7,1]     &   1     & 2.05   & 1     & 2.05   & 1     & 2.05   \\
      2 &(0.65,0.7]  &   1     & 2.37   & 1     & 2.37   & 0     & 0      \\
      3 &(0.6,0.65]  &   0     & 0      & 1     & 2.7778 & 0     & 0      \\
      4 &(0.5,0.6]   & 0       & 0      & 1     & 4      & 0     & 0    \\
      5 &(0.4,0.5]   & 1/4     & 1.5625 & 1/4   & 1.5625 & 1/4   & 1.5625\\
      6 & (0.35,0.4] & 0.26    & 2.123  & 0.26  & 2.123  & 0.26  & 2.123  \\
      7 &(1/3,0.35]  & 0.8/3   & 2.4    & 0.8/3 & 2.4    & 0.2/3 & 0.6   \\
      8 &(0.3,1/3]   &  1/9    & 1.235  &  1/9  & 1.235  &  1/9  &  1.235\\
      9 &(1/4,0.3]   & 0.1338  & 2.141  & 0.0829 & 1.327 & 0.1338 & 2.141 \\
 10..17 &(1/5,1/4]   &1.99x^2  & 1.99   &1.235x^2& 1.235 &1.99x^2 & 1.99  \\
%%      11 &(1/6,1/5]   & 0.05067 & 1.824  & 0.034 & 1.224  & 0.05067 & 1.824 \\
%%      12 &(1/7,1/6]   &0.03410  & 1.6705 & 0.025 & 1.225 &0.03410 & 1.6705 \\  
%%      13 &(1/8,1/7]   & 0.02606 & 1.6679 & 0.01837 & 1.1756 & 0.02606 & 1.6679 \\
%%      14 &(1/9,1/8]   & 0.02073 & 1.6791 & 0.9/64 &1.2& 0.02073  & 1.6791 \\   
%%      15 &(0.1,1/9]   & 0.017   & 1.7    &  0.1/9  & 1.2 & 0.017   & 1.7  \\  
%%      16 &(1/11,0.1]  & 0.01089 & 1.3176 & 0.0095 & 1.2 & 0.01089  & 1.3176 \\ 
%%      17 &(0,1/11]    &1.2x^2   & 1.2   &1.2x^2  & 1.2   &1.2x^2  & 1.2 \\   
  \hline
 \end{array}
 \end{displaymath}

\paragraph{ Case 3:} $ W_3(X) \le   2.12$.

 \noindent If $m_1+ m_2 + m_3 + m_4 = 0$, i.e., 
 no  type $1,2,3,4$ items in $X$,
 then $m_6 + m_7 \le 4$,
 \[
  W_3(X) =   W_{3,2}(X)  \le 0.26m_6 + \frac{0.8m_7}{3} +
   1.5625(1-0.35^2m_6 -\frac{m_7}{9})
   < 2.
 \]  
 Else $m_1 =1$ then $m_i=0$, where $2 \le i \le 8$,
\[
  W_3(X) =   W_{3,2}(X)  \le 2.05.
 \]  
 Else $m_2 = 1$. Then no type $1,3,4,5,6$ items  in $X$,
 $m_7 + m_9 \le 5$ and $m_7 \le 3$.
 \[
  W_3(X) = W_{3,2}(X) \le 1 + \frac{0.8m_7}{3} + 0.0829m_9 + 1.235(1-0.65^2 - m_7/9 -m_9/16) < 2.12. 
 \]  
 Else  $m_3 + m_4 = 1$.
 Then no type $1,2,3$ items  in $X$.
  \[
  W_3(X) \le W_{3,1}(X) \le 0 + 2.4 (1 - 0.5^2)  =  1.8.
 \]  

\paragraph{ Case 4:} $  W_4(X) = \sum_{p \in X} E_{4,1}(p) s(p)^2
     \le 2.141    \sum_{p \in X} s(p)^2 \le 2.141$.

So,  $\mathcal{P}(W) \le 2.1439$.
\end{proof}

\begin{theorem}
 The asymptotic performance ratio of $\mathcal{A}$ for cube  packing is at most 2.6852.
\end{theorem}
\begin{proof}
 For cube packing, we set parameters $\alpha_i$ and $\theta_i$ in the following table.
 \begin{displaymath}
 \begin{array}{|c|l|l|l|l|l|l|l|l|l|l|}
 \hline
 i           & 1-4 & 5 & 6 & 7 & 8 & 9 & 10 & 11 & 12-16 \\
 \hline
 \alpha_i  & 0 & 0 & 0.12 & 0.2 & 0 & 0.325 & 0.2096 & 0.15& 0 \\
 \hline
 \theta_i & 0 & 0 & 7 &  7 & 0 & 19 & 37 & 61 &   0    \\
 \hline
 \beta_i^3 & 1 & 8 & 8 & 8 & 27 & 27 & 64 & 125 & (i-6)^3 \\
 \hline
 \end{array}
\end{displaymath}
 Here we set  $\alpha_i=0$ for $12 \le i \le 16$.
 So, their weights are defined as $1/\beta_i^3$.

%%  Given an item $p$ with size $x$,
%%   an efficient function  $E_i(p)$ is defined  as $W_i(p)/x^3$,
%%  then we have the following table.
 We first give two tables and then use them to calculate  $\mathcal{P}(W)$.
 \begin{displaymath}
 \begin{array}{|c|l|l|l|l|l|l|l|}
 \hline
  i & (t_{i+1},t_i] & W_{1,1}(p) & E_{1,1}(p)  & W_{2,1}(p) & E_{2,1}(p) &W_{2,2}(p) & E_{2,2}(p) \\
 \hline
      1 &(0.7,1]    & 1     & 2.9155 &   1     & 2.9155 & 1      & 2.9155    \\
      2 &(0.65,0.7] & 1     & 3.65   &   1     & 3.65   & 1      & 3.65     \\
      3 &(0.6,0.65] & 1     & 4.63   &   1     &  4.63  & 1      &  4.63     \\
      4 &(0.5,0.6]  & 1     & 8      & 0       & 0      &    1   & 8     \\
      5 &(0.4,0.5]  & 1/8   & 1.9532 & 1/8     & 1.9532 & 1/8    & 1.9532 \\
      6 & (0.35,0.4]& 0.11  & 2.5656 & 0.1272  & 2.966  & 0.1272 &  2.966  \\
      7 &(1/3,0.35] & 0.1   & 2.7    & 0.1286  & 3.472  &   0.1  & 2.7  \\
      8 &(0.3,1/3]  & 1/27  & 1.372  &  1/27   & 1.372  &  1/27  & 1.372 \\
      9 &(1/4,0.3]  & 0.025 & 1.6    &0.04211  & 2.6948 & 0.025  & 1.6    \\
      10 &(1/5,1/4] & 0.0124& 1.55   & 0.01802 & 2.252  & 0.0124 & 1.55    \\
      11 &(1/6,1/5] & 0.0068& 1.4688 & 0.0093  & 2      & 0.0068 & 1.4688  \\
%%      12  &(1/7,1/6]& 1/216 & 1.59 & 1/216 & 1.59  & 1/216 & 1.59  & 1/216 & 1.59  \\      
%%      13 &(1/8,1/7]& 1/7^3 & 1.5   & 1/7^3 & 1.5   & 1/7^3 & 1.5   & 1/7^3 & 1.5   \\  
%%      14 &(1/9,1/8]& 1/8^3 & 1.43 & 1/8^3 & 1.43 & 1/8^3 & 1.43  & 1/8^3 & 1.43 \\   
%%      15 &(0.1,1/9]& 1/9^3 & 1.38 & 1/9^3 & 1.38 & 1/9^3 & 1.38  & 1/9^3 & 1.38   \\        
%%      16 &(1/11,0.1]& 0.001 & 1.331& 0.001 & 1.331 & 0.001 & 1.331  & 0.001 & 1.331  \\  
     12..17 &(0,1/6]& 1.59x^3 & 1.59 & 1.59x^3 & 1.59 & 1.59x^3 & 1.59  \\
 \hline
 \end{array}
 \end{displaymath}

\paragraph{ Case 1:} $ W_1(X) \le   2.6852$.
 
 \noindent If $m_1+m_2 + m_3 + m_4 = 0$, i.e., no  type $1,2,3,4$ items in $X$,
 then $m_6 + m_7 \le 8$,
\[
    W_1(X) \le 0.11m_6+0.1m_7 + 1.96(1-0.35^3m_6-m_7/27)
   \le 2.3.
 \]  
 Else $m_1 = 1$. Then  $m_i=0$, where $2 \le i \le 8$,
\[ 
W_1(X) \le 1 + 1.6(1-0.7^3 ) = 2.0512.
\] 
 Else $m_2 = 1$. Then  no  type $1,3,4,5,6$ items  in $X$ and $m_7  \le 7$,
 \[
  W_1(X) \le 1 + 0.1\times 7  + 1.6(1-0.65^3 - 7/27)
   \le 2.546.
\]
 Else $m_3 =1$. Then no  type $1,2,4,5$ items  in $X$ and $m_6 + m_7 \le 7$, 
 \[
     W_1(X)\le 1+ 0.11 m_6 + 0.1m_7 + 1.6(1-0.6^3 -0.35^3m_6-m_7/27) \le 2.5646.
 \]
 Else $m_4 = 1$. Then $m_1 + m_2 + m_3 = 0$ and $m_5 + m_6 + m_7 \le 7$,
 \[
  \begin{array}{lll}
  W_1(X) & \le &  1+  m_5/8 + 0.11 m_6 + 0.1m_7 \\
                  & & + 1.6(1-0.5^3 - 0.4^3m_5 -0.35^3m_6-m_7/27) \\
                  & < &  2.6852.
   \end{array}
 \]
 The last inequality follows from $m_7=7$ and $m_5=m_6=0$.

\paragraph{ Case 2:} $ W_2(X) \le   2.6646$.

 \noindent If $m_1 +m_2 + m_3 + m_4 = 0$, i.e., 
 no  type $1,2,3,4$ items in $X$,  then  $m_6 + m_7 \le 8$,
\[
    W_2(X) = W_{2,2} \le 0.1272m_6+0.1m_7 + 1.96(1-0.35^3m_6-m_7/27)
   \le 2.4.
 \]  
Else $m_1 + m_2 = 1$ . Then no type $1,4,5,6$ items  in $X$,
\[
  W_2(X) = W_{2,2}(X) = W_1(X) \le 2.546.
\]
 Else  $m_3 = 1$. Then no  type $1,2,4,5$ items in $X$ and $m_6 + m_7 \le 7$, 
 \[
     W_2(X)\le 1+ 0.1272 m_6 + 0.1m_7 + 1.6(1-0.6^3 -0.35^3m_6-m_7/27) \le 2.6646.
 \]
Else $m_4 =1$. Then no type $1,2,3$ items  in $X$ and $m_6 + m_7 \le 7$,
  \[
  W_2(X) \le W_{2,1}(X) \le 0 + 0.1272m_6 + 0.1286 m_7 + 
           2.6948 (1 -0.35^3m_6-m_7/27 - 1/8) < 2.5595.
 \]  
 The last inequality  holds for $m_6 = 0$ and $m_7 = 7$.

 \begin{displaymath}
 \begin{array}{|c|l|l|l|l|l|l|l|}
 \hline
  i & (t_{i+1},t_i] &  W_{3,1}(p) & E_{3,1}(p)  & W_{3,2}(p) & E_{3,2}(p)  & W_{4,1}(p) & E_{4,1}(p) \\
 \hline
      1  &(0.7,1]   &   1     & 2.9155 & 1      & 2.9155 & 1      & 2.9155   \\
      2  &(0.65,0.7]&   1     & 3.65   & 1      & 3.65   & 0      & 0     \\
      3  &(0.6,0.65]& 0       & 0      &   1    &  4.63  & 0      & 0      \\
      4  &(0.5,0.6] & 0       & 0      & 1      & 8      & 0      & 0   \\
      5  &(0.4,0.5] & 1/8     & 1.9532 & 1/8    & 1.9532 & 1/8    & 1.9532 \\
      6  &(0.35,0.4]& 0.1272  & 2.966  & 0.1272 &  2.966 & 0.1272 &2.966  \\
      7  &(1/3,0.35]& 0.1286  & 3.472  & 0.1286 & 3.472  & 0.03   & 0.81 \\
      8  &(0.3,1/3] &  1/27   & 1.372  &  1/27  & 1.372  &  1/27  & 1.372 \\
      9  &(1/4,0.3] & 0.04211 & 2.6948 & 0.025  & 1.6    & 0.04211& 2.6948   \\
      10 &(1/5,1/4] & 0.01802 & 2.252  & 0.0124 & 1.55   &0.01802 & 2.252  \\
      11 &(1/6,1/5] & 0.0093  & 2      & 0.0068 & 1.4688 & 0.0093 & 2  \\
%%      12  &(1/7,1/6]& 1/216 & 1.59 & 1/216 & 1.59  & 1/216 & 1.59  & 1/216 & 1.59  \\      
%%      13 &(1/8,1/7]& 1/7^3 & 1.5   & 1/7^3 & 1.5   & 1/7^3 & 1.5   & 1/7^3 & 1.5   \\  
%%      14 &(1/9,1/8]& 1/8^3 & 1.43 & 1/8^3 & 1.43 & 1/8^3 & 1.43  & 1/8^3 & 1.43 \\   
%%      15 &(0.1,1/9]& 1/9^3 & 1.38 & 1/9^3 & 1.38 & 1/9^3 & 1.38  & 1/9^3 & 1.38   \\        
%%      16 &(1/11,0.1]& 0.001 & 1.331& 0.001 & 1.331 & 0.001 & 1.331  & 0.001 & 1.331  \\  
     12..17 &(0,1/6]& 1.59x^3 & 1.59 & 1.59x^3 & 1.59 & 1.59x^3 & 1.59  \\
 \hline
 \end{array}
 \end{displaymath}
 
\paragraph{ Case 3:} $ W_3(X) \le   2.646$.

 \noindent If $m_1+ m_2 + m_3 + m_4 = 0$, i.e., 
 no  type $1,2,3,4$ items in $X$,
 then  $m_6 + m_7 \le 8$,
\[
    W_3(X) = W_{3,2} \le 0.1272m_6+0.1286m_7 + 1.96(1-0.35^3m_6-m_7/27)
   \le 2.41.
 \]  
Else $m_1 =1$. Then $m_i=0$, where $2 \le i \le 8$,
\[
  W_3(X) = W_{3,2}(X) = W_1(X) \le 2.0512.
\]

 Else $m_2 = 1$. Then  no  type $1,3,4,5,6$ items  in $X$ and $m_7  \le 7$,
 \[
  W_3(X) = W_{3,2}(X) \le 1 + 0.1286\times 7  + 1.6(1-0.65^3 - 7/27)
   \le 2.646.
\]

 Else  $m_3 + m_4 = 1$.
 Then no type $1,2$ items  in $X$.
  \[
  W_3(X) \le W_{3,1}(X) = W_{2,1}(X) \le 2.5595.
 \]  

\paragraph{ Case 4:} $  W_4(X) \le  2.63$.

\noindent If $m_1+ m_2 + m_3 + m_4 = 0$, i.e., 
 no  type $1,2,3,4$ items in $X$,  $m_6 + m_9 \le 27$ and $m_6 \le 8$,
\[
   W_4(X) \le 0.1272m_6+0.04211m_9 + 2.252(1-0.35^3m_6-m_9/64)
   \le 2.63.
 \]  
%% The last inequality  holds for $m_6 = 8$ and $m_9 = 19$.

Else $m_1 =1$. Then $m_i=0$, where $2 \le i \le 8$.
 And $m_9 \le 19$, 
\[ 
 W_4(X) \le 1 + 19 \times 0.04211 + 2.252(1-0.7^3 -19/64)
 \le 2.62.
\] 
Else $m_2 + m_3 + m_4 = 1$. 
 Then  $m_6 \le 7$,
  \[
  W_4(X) \le 0 + 0.1272 m_6  + 
           2.6948 (1 -0.35^3m_6 - 1/8) < 2.4396.
 \]

 So, $\mathcal{P}(W) < 2.6852$.
\end{proof}
\section{Concluding Remarks}
In this page, we reduce the gaps  between the upper and lower bounds of
online square packing and cube packing.
But the gaps are still large.
It seems  possible to use computer proof as the one  in \cite{s02}
to get  a more precise upper bound.
But, the analysis becomes more complicated and more difficult than 
the one in \cite{s02},
since we are faced to solve a two dimensional knapsack problem, rather
than one dimensional knapsack problem \cite{s02}.
So, how to reduce the gaps is a challenging open problem.

%% \normalsize
%% \newpage
%% \begin{center}
%% {\bf Appendix}
%% \end{center}

%% \begin{table} [h]

%% \caption{ $W_{1,1}(p)$, $W_{2,1}(p)$, $W_{2,2}(p)$}
%% \label{table:t1}
%% \end{table}

%% \begin{table} [ht]

%% \caption{  $W_{3,1}(p)$, $W_{3,2}(p)$ and $W_{4,1}(p)$}
%% \label{table:t2}
%% \end{table}

\end{document}